\documentclass[12pt]{article}
\usepackage{times}
\usepackage{geometry}
\usepackage[utf8]{inputenc}
\usepackage{enumitem,amssymb}
\usepackage{ragged2e}
\newlist{thematic}{itemize}{8}
\setlist[thematic]{label=$\square$}
\usepackage{pifont}
\usepackage{wasysym}
\usepackage{graphicx}
\usepackage[font=footnotesize,labelfont=bf]{caption}
\usepackage{subcaption}
\usepackage{url}
\usepackage{wrapfig}

\begin{document}

\pagenumbering{gobble}

{\raggedright
\huge
Astro2020 Science White Paper \linebreak

Gravitational wave astronomy with LIGO and similar detectors in the next decade \linebreak
\normalsize

\noindent \textbf{Thematic Areas:} \hspace*{60pt} $\square$ Planetary Systems \hspace*{10pt} $\square$ Star and Planet Formation \hspace*{20pt}\linebreak
$\XBox$ Formation and Evolution of Compact Objects \hspace*{31pt} $\XBox$ Cosmology and Fundamental Physics \linebreak
  $\square$  Stars and Stellar Evolution \hspace*{1pt} $\square$ Resolved Stellar Populations and their Environments \hspace*{40pt} \linebreak
  $\square$    Galaxy Evolution   \hspace*{45pt} $\XBox$             Multi-Messenger Astronomy and Astrophysics \hspace*{65pt} \linebreak
  
\textbf{Principal Author:}

Name:	David Shoemaker, LIGO Scientific Collaboration spokesperson
 \linebreak						
Institution:  Massachusetts Institute of Technology
 \linebreak
Email: david.shoemaker@ligo.org
 \linebreak
Phone:  (617) 253 6411
 \linebreak
 
\textbf{Co-authors:} (names and institutions)
LIGO Scientific Collaboration; complete author list in https://dcc.ligo.org/LIGO-M1800241
  \linebreak

\textbf{Abstract :}

We describe the plans for gravitational-wave observations and astrophysics that will be 
carried out by the LIGO Scientific Collaboration (LSC) in the next decade using data from the LIGO Observatories in the US, and sister facilities abroad in Europe, Japan and India. We provide an overview of gravitational wave signal types that we are targeting, and the role of gravitational waves in time-domain multi-messenger astronomy.  We briefly summarize what we can infer from the properties of detected signals, including astrophysical event rates and populations, tests of gravitational-wave properties, highly-dynamical and strong-field tests of General Relativity, probing matter under extreme conditions in neutron stars, and making cosmological measurements with gravitational-wave sources.
}
\
\thispagestyle{empty}

\pagebreak

\pagenumbering{arabic}

\noindent {\Large \bf Introduction}
\smallskip

The past three years have witnessed the birth of observational gravitational-wave astronomy, starting 
with the first detection 
of a binary black hole merger on September 14 2015 \cite{Abbott:2016blz}, followed by 
discoveries of nine more in the first and second LIGO/Virgo Observing runs \cite{LIGOScientific:2018mvr}, and the spectacular multi-messenger 
observation of a merger of neutron stars on August 17, 2017 \cite{Monitor:2017mdv, GBM:2017lvd}. 

These detections were enabled by a nearly three decade long effort to build  
Advanced LIGO 
~\cite{aLIGO} comprising two laser interferometric gravitational-wave detectors with suspended mirrors,  
laser beams traveling in vacuum through 4 km perpendicular arms in each detector, 
to detect sub-nuclear distance scale changes in distance.  
The LIGO Scientific Collaboration (LSC) works closely with the Virgo and KAGRA collaborations operating gravitational-wave detectors
in Europe and Japan to ensure 
coordinated observations by the global network.
\begin{figure}[htbp] 
\begin{minipage}[t]{0.5\textwidth}
   \centering
   \includegraphics[width=3in]{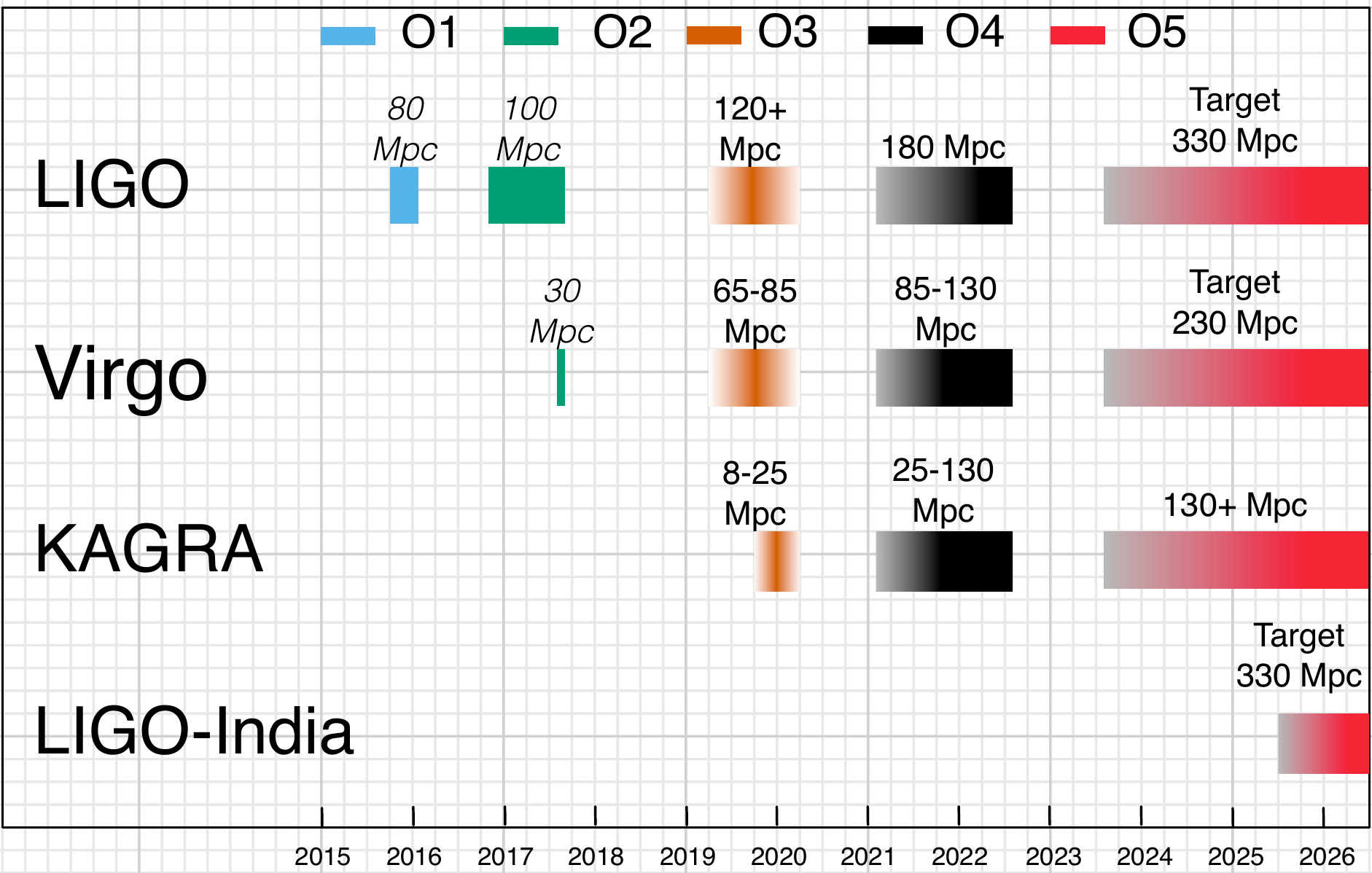} 
   \caption{Planned sensitivity evolution and observing runs of ground-based detectors. Numbers in Mpc indicate average reach to binary neutron star mergers. }
   \label{fig:runs}
   \end{minipage}
   \hspace{0.1in}
\begin{minipage}[t]{0.5\textwidth}
   \centering
   \includegraphics[width=3.3in]{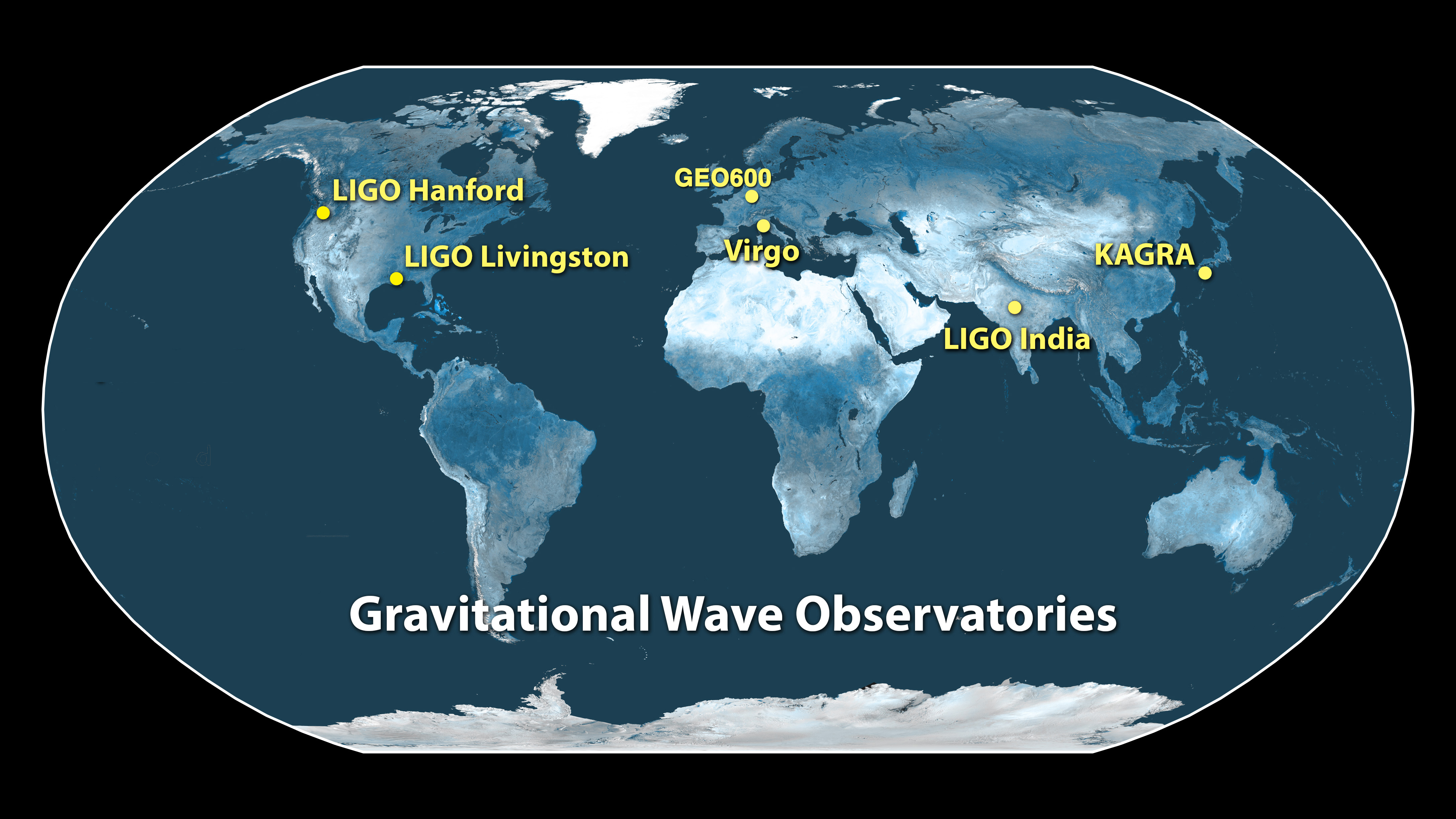} 
   \caption{A map of the global ground-based gravitational-wave detector network.}
   \label{fig:map}
   \end{minipage}
   \vspace{-10pt}
\end{figure}
In this white paper, we describe plans for gravitational-wave observing campaigns and expected science goals in the coming decade.
As shown in Fig.~\ref{fig:runs}, the Advanced LIGO detectors took data between September 2015 and January 2016 in their first Observing run (O1), and then again with improved sensitivity in O2, between November 2016 and August 2017.  The  Virgo detector  
\cite{Virgo:2014hva} joined O2 on August 1st 2017,  providing  greatly improved sky localization  
of the detected events. The improved localization and rapid alerts 
led to  the detection of an electromagnetic counterpart to the binary neutron star merger \cite{Monitor:2017mdv, GBM:2017lvd}. This counterpart, 
spanning all bands of the electromagnetic spectrum, allowed the first direct association between a binary neutron star merger and a 
short gamma-ray burst, and the first unambiguous identification of a kilonova. 

In the next few years, the Advanced LIGO and Virgo detectors will continue to observe and analyze data together, 
and are expected to reach the sensitivity to
which they were designed \cite{Abbott2018:LivRev}. KAGRA \cite{KAGRA}
is expected to join in the gravitational wave network in 2019.  The GEO600 \cite{Luck:2010rt}
detector will provide coverage for exceptional 
events during times when no other detectors will be operating, and will otherwise concentrate on testing technologies for future detectors \cite{Affeldt_2014}.
The greatest scientific return is possible when all operating GW detectors combine their data. 

Funding from the US, UK and Australia has been 
secured for the ``A+" 
detector upgrade \cite{NSFAplus}, implementing further
sensitivity improvements beyond the current Advanced LIGO design. 
Toward the middle of the next decade, 
a new observatory in India \cite{LIGO-India} will host an Advanced LIGO detector to further enhance the network sensitivity. 
This decade will see an improvement of a factor of several in astrophysical distance reach, as well as a significant increase in
observing time, which translates into a factor of a hundred increase in rate of detections of black holes and neutron stars. However, detection of mergers of more massive black hole at the centers of galaxies will need other instruments, like space detectors and pulsar timing arrays; detections of neutron star mergers at cosmological distances will need a new generation of ground based facilities. 

The LSC is  working to bring its advanced detectors to their design sensitivity,  improve their astrophysical reach with technological advances, and undertake observing runs collecting calibrated gravitational-wave data. 
It will perform searches for gravitational waves, some in near real time, and low-latency public alerts will be issued to the broader astronomical community to enable
multi-messenger observations of gravitational-wave events. The LSC extracts the details of the gravitational-wave signals
from the data and, using the measured properties of the signal, presents in publications the astrophysical implications.
Following a proprietary period, the LIGO data are made public, enabling other scientists to independently search the data~\cite{LIGODMP}.  More details of the LSC's Scientific Goals are available in the LSC Program \cite{LSCprogram}
and accompanying White Papers.

\bigskip
\noindent {\Large \bf Gravitational Wave Targets}
\smallskip

We list in this section the gravitational wave targets that the LSC and Virgo will be searching for
in the near term, with KAGRA and LIGO-India joining when possible. 

\begin{wrapfigure}{r}{0.5\textwidth}
\vspace{-25pt}
\begin{center}
   \includegraphics[width=3in]{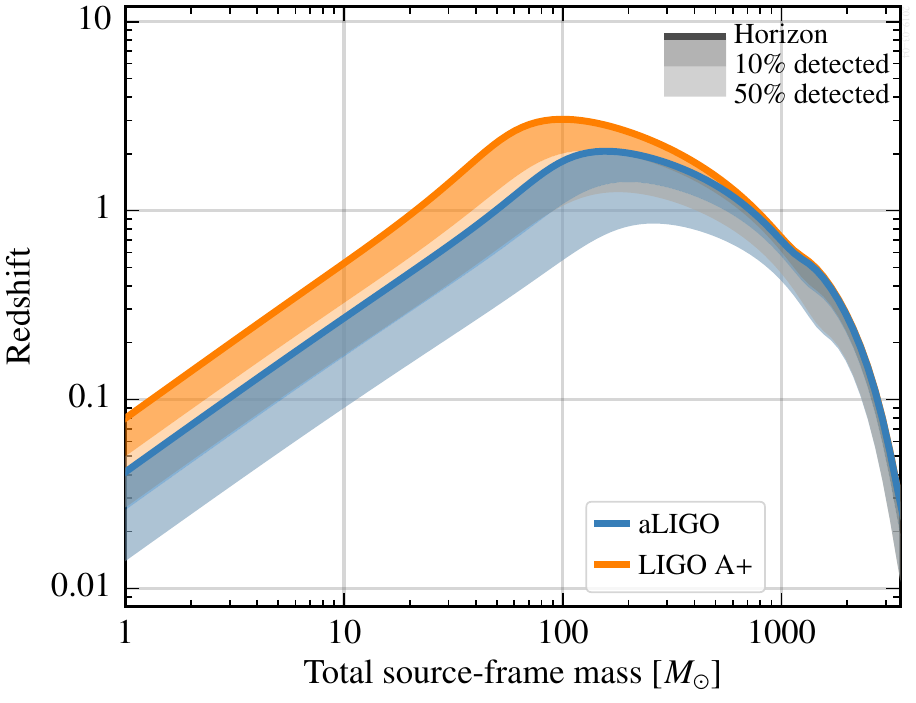} 
   \end{center}
   \vspace{-15pt}
   \caption{Astrophysical reach of the aLIGO and A+ detectors as a function of total mass of merging binary systems.
   Contours show the maximum reach (Horizon) and distance at which 10\% and 50\% of sources are observable.}
   \label{fig:cbc_range}
   \vspace{-15pt}
\end{wrapfigure}
\textbf{Gravitational waves emitted during the coalescence of compact binaries}. We will search for mergers of compact binaries
that produce gravitational waves in the sensitive frequency range of the LIGO detectors, most prominently binary systems with neutron stars and stellar-mass black holes. These searches  are run in low latency to provide alerts to electromagnetic observers. 

Binary black hole mergers are expected to be the dominant class of gravitational-wave observations, with A+ having sensitivity
to systems with masses up to $1000 M_\odot$ and out to redshifts of $z=3$, as shown in figure \ref{fig:cbc_range}.
The frequency of binary-black-hole observations is expected to increase to as much as several detections per week by
the end of the decade, and we will provide a comprehensive catalog of binary black hole populations during the decade. 
 
Binary neutron stars can be observed to $z\sim 0.1$.  With only one detection to date, the astrophysical rate of binary neutron 
star mergers is not yet well constrained.  However, several detections per year are likely.
We also search for neutron-star--black-hole binaries, with observations expected within the decade \cite{Abbott2018:LivRev}.

\textbf{Searches for unmodeled transient gravitational wave signals.} We will search for transients with durations from a few milliseconds up
to hours or days \cite{Abbott:2016ezn, Abbott:2017muc}
. Expected sources include core-collapse supernovae, soft gamma repeaters, neutron star glitches,
proto-neutron stars and accretion disks, and cosmic string cusps, as well as high mass strong black hole mergers and eccentric binary systems. 
Blind, all-sky searches will also allow the discovery of previously unknown sources. 
Searches for short transients will be run in low latency. 

\textbf{Gravitational waves associated with known astronomical transients.} We will search for transient gravitational wave 
signals around known electromagnetic transients such as gamma-ray bursts \cite{Abbott:2016cjt}, fast radio bursts, 
supernovae \cite{Abbott:2016tdt}, and magnetar flares. Results from multi-messenger observations of these sources will provide insight on their structure and origin. By using the known times and sky locations of these electromagnetic transients and, where applicable, the expected gravitational wave signals, we will perform targeted gravitational wave searches with improved sensitivity over blind all-sky searches. Some of these searches will be performed in low-latency mode to allow for alerts to be issued to the broader community.

\textbf{Gravitational waves emitted by previously unknown non-axisymmetric neutron stars.} 
Neutron stars are invaluable laboratories for physics under extreme conditions. 
We will search for continuous gravitational wave emission from fast-spinning galactic 
neutron stars \cite{Pisarski:2019vxw}, both isolated and in binary systems. These stars can emit gravitational 
radiation through a variety of mechanisms, including rotation with elastic deformations, 
magnetic deformations, unstable r-mode oscillations, and free precession, all of which 
operate differently in accreting and non-accreting stars. With longer analysis times 
and improved detector sensitivity, it is likely that new sources in the Milky Way will 
be localized, although there is still a large uncertainty on possible strength of 
gravitational wave emission.  

\textbf{Continuous gravitational waves emitted by known pulsars and other 
promising sources.} 
We will search in greater depth for continuous gravitational waves from known sources 
\cite{Abbott:2019bed}
exploiting astrophysical measurements, such as the frequency evolution 
of known pulsars and/or the locations of other promising sources, like 
supernovae and known X-ray binaries \cite{Abbott:2018qee}.  A new generation of telescopes will become 
available during the next decade, which will increase by an order of magnitude 
the number of known and monitored pulsars. Improvements in detector sensitivity, 
longer duration observing runs, new electromagnetic  observations, and algorithmic 
improvements will all contribute to a compelling possibility for detection and
 observation of continuous gravitational waves. The detection of gravitational 
 radiation from these sources will facilitate decisive progress on one of the most 
 fundamental questions of modern science: the composition and state of matter at 
 extreme densities. It will also allow testing deviations from general relativity, 
 for example, wave polarization or orbital evolution different from that predicted by general relativity.

\textbf{Searches for astrophysical and cosmological gravitational wave backgrounds.} 
We will search for stochastic gravitational-wave backgrounds,
formed from the superposition of signals that are individually either too weak or
too numerous to resolve \cite{LIGOScientific:2019vic}.  The primary target will be the
background from unresolved binary mergers of 
stellar-mass black holes and/or neutron stars throughout the universe.
Rate estimates 
suggest that this background may be detectable by 2024
\cite{TheLIGOScientific:2016wyq},
but perhaps earlier using advanced analysis methods. 
The detection
of this background will provide information about stellar-mass
binary-black-hole populations at much larger distances than those
accessible for the resolved mergers.  It will also complement searches
for gravitational-wave backgrounds at much lower frequencies, 
complementing likely detections before the end of the decade by pulsar-timing array and possibly by  CMB searches. 

We will also search for gravitational-wave backgrounds produced by
cosmic string networks, and other backgrounds of cosmological origin
\cite{TheLIGOScientific:2016dpb}.
We will search for anisotropies in the background, where the
anisotropy may be correlated with large-scale structure, and
deviations from general relativity, allowing for non-standard
polarizations of the component gravitational waves \cite{Abbott:2018utx}.
After the initial detections, it will be important to
extract the properties of the sources that give rise to these
backgrounds.

\bigskip
\noindent {\Large \bf Gravitational Wave Astronomy}
\smallskip

The following list describes the measurements to be carried out for gravitational wave detections and potential conclusions to be drawn from non-detections.

\textbf{Signal Characterization.} We will extract the physical properties of the observed gravitational wave signals. For events where the source is well modeled, such as a binary merger, we will extract the physical parameters of the source
\cite{TheLIGOScientific:2016wfe}. Where the signal morphology is not well modeled, such as a core-collapse supernova, we will reconstruct the waveforms. Where possible, we will determine best-fit maps of sky position and distance.
For binary mergers, the estimated masses, spin angular momenta, orbital eccentricities and distances will shed light on their (poorly understood) astrophysical formation mechanism \cite{TheLIGOScientific:2016htt}. The tidal deformation of neutron stars will contain imprints of the dense nuclear equation of state \cite{Abbott:2018exr}. Binary merger signals will potentially contain interesting effects such as spin precession and higher order modes. Detection of galactic supernova signals will allow us to probe the core-collapse supernova mechanism while detection of the astrophysical stochastic gravitational-wave background will constrain the star formation history. 

\textbf{Public alerts.} Starting with the third Observing run (O3) in 2019, LIGO/Virgo alerts 
will be public. We will issue prompt and open public alerts of significant gravitational wave events in O3 to allow for follow-up 
observations with electromagnetic and neutrino observatories. Alerts will be distributed for compact binary coalescence (CBC) events 
including binary black hole, binary neutron star, and neutron star black hole systems as well as unmodeled transient 
gravitational wave signals. We maintain the \emph{LIGO/Virgo Public Alerts User Guide} \cite{alerts} for both professional astronomers and science enthusiasts 
who are interested in receiving alerts and real-time data products related to gravitational wave events. Alerts will be distributed through 
NASA's Gamma-ray Coordinates Network (GCN). The user guide provides a brief overview of the procedures for vetting and sending alerts, describes 
their contents and format, and includes instructions and sample code for receiving GCN Notices and decoding sky maps.

\textbf{Astrophysical rates and populations.} We will use the observed individual events, primarily compact binary 
coalescences of black holes and neutron stars, to determine the underlying population of sources in the universe, 
taking into account selection effects. For binary mergers, this will enable inferences of the distribution of binary masses, including
any mass gaps, spin magnitudes and orientations \cite{TheLIGOScientific:2016pea, LIGOScientific:2018jsj}.  We will interpret the detected populations in terms of existing models of compact binary formation and evolution \cite{TheLIGOScientific:2016htt}.
When other sources are observed, we will similarly constrain the underlying astrophysical populations, and where no sources
are observed, rate upper limits will be used to constrain the astrophysical population properties.
We will determine the implications of stochastic background search results for various cosmological and astrophysical models,
including models based on cosmic string cusps and kinks, inflationary scenarios and models due to mergers of binary neutron stars
and/or black holes.

\textbf{Tests of gravitational-wave properties.} In General
  Relativity (GR), gravitational waves propagate at a constant speed,
  independent of frequency, equal to the speed of
  light. Gravitational wave observations, both with and without
  electromagnetic counterparts, can be used to look for variations of
  the speed of gravity (either from the speed of light or as a
  function of gravitational wave frequency), and Lorentz and parity violations \cite{Monitor:2017mdv, TheLIGOScientific:2016src}. 
Observations of
  gravitational wave transients or stochastic gravitational wave
  backgrounds in a network of detectors \cite{Abbott:2017oio}, or of continuous
  gravitational waves in one or more detectors, allow us to probe the
  polarization content of the signal and look for the existence of
  more polarizations than predicted by GR. 

\textbf{Highly-dynamical and strong-field tests of General
    Relativity.} Gravitational wave observations from binary
  coalescences and inference of source properties are made through
  waveform models built solving the Einstein's field
  equations. Waveform models can also be built in theories of gravity
  alternative to GR, where gravitational interactions are mediated by
  extra degrees of freedom --- for example scalar fields. Thus,
  gravitational wave observations allow us to test for deviations from
  GR~\cite{TheLIGOScientific:2016pea, TheLIGOScientific:2016src, Abbott:2018lct}, 
  including violations of the strong equivalence principle. These
  violations may occur during the early inspiral, as deviations from
  the post-Newtonian coefficients computed in GR, and also during the
  merger and ringdown, where the signal may differ from GR predictions
  in numerical relativity simulations, and from quasi normal modes of
  black holes in GR. The latter may occur if black hole's uniqueness
  theorems are violated or the compact object has a ``surface''
  instead of an horizon. We will search for these possible deviations,
  both in individual signals and by coherently analyzing the
  population of observed signals.

\textbf{Probing extremes of matter.} 
The detection of continuous gravitational waves from neutron stars, and further observations of
neutron star mergers, will give us insight into the 
ultra-dense interiors of neutron stars. Observations of continuous waves will provide clues about neutron star 
formation or accretion and magnetosphere physics (from observation of binaries), their spin, 
thermal and magnetic field evolution, the nature of cold dense matter, 
and phase transitions in QCD. Electromagnetic observations of the star could be especially
 helpful in establishing distance, and hence absolute signal strength, and in relating potential electromagnetic 
 pulse phase to gravitational
 wave signal phase, which is relevant to interpreting the neutron star non-axisymmetry.
 
The many expected observations of neutron stars in binary mergers, especially if we extract details of the inspiral and the post-merger waveforms, will allow us to probe the neutron-star equation of state~\cite{Abbott:2018exr, Abbott:2018wiz}. 
Since the coalescence of binary systems involving neutron stars produces electromagnetic waves, combining electromagnetic and gravitational wave observations can also yield insight into the mechanisms for prompt and post-merger electromagnetic 
emission \cite{GBM:2017lvd}. 
 In the fortunate event of a nearby supernova [probability of O(10\%) by 2030], combining neutrino and gravitational wave
 observations can yield insight into the collapse and explosion mechanisms.  
  
\textbf{Gravitational wave cosmology.} The gravitational waveform emitted during a binary merger can be used to
obtain a measurement of the luminosity distance to the binary, known as the standard siren method.  Thus, given an accurate measurement of the source redshift it is possible to probe the expansion history of the universe and, concretely, to measure the Hubble constant \cite{Abbott:2017xzu}. The redshift measurement can either be from an
electromagnetic observation, directly from the properties of the gravitational wave signal ({\it e.g.}, merger physics in neutron star
mergers) or statistically derived from overlaying a galaxy catalog on the source localization. We expect that the many detections  in the coming decade will lead to an era of precision gravitational-wave cosmology.

\bigskip
\noindent {\Large \bf Outlook}
\smallskip

We have described the exciting science that will be carried out in the next decade using data from a ground-based gravitational wave observatory network including the LIGO Observatories and sister facilities abroad in Europe with improving sensitivities, and then with a larger network including facilities in Japan and India. From the properties of many signals to be detected with the network, we will infer astrophysical event rates and populations, perform tests of gravitational-wave properties, highly-dynamical and strong-field tests of General Relativity, probe matter under extreme conditions in neutron stars, make cosmological measurements with gravitational-wave sources - and probably have new astronomical milestone discoveries. 

\pagebreak

\bibliographystyle{unsrt}
\bibliography{lscwp.bib}

\end{document}